\documentclass{aip-cp}

\usepackage[numbers]{natbib}
\usepackage{rotating}
\usepackage{graphicx}
\begin{document}

% Title portion
\title{Equation of State Model for the $\gamma-\alpha$ Transition in Ce}

\author[aff1]{C. W. Greeff}
%\eaddress[url]{http://www.aip.org}
\author[aff1]{S. D. Crockett}
%\eaddress{anotherauthor@thisaddress.yyy}
\author[aff1]{K. G. Honnell}
%\eaddress{anotherauthor@thisaddress.yyy}

\affil[aff1]{ Los Alamos National Laboratory, Los Alamos, NM 87545}
%\corresp[cor1]{Corresponding author: your@emailaddress.xxx}

\maketitle

\begin{abstract}
The element Ce exhibits an isostructural $\gamma-\alpha$ phase transition 
with a 15\% volume change at room temperature. The phase boundary
ends in a critical point at 1.5 GPa and 480 K.
We describe a model for the equation of state of Ce in the $\gamma-\alpha$
transition region. The model is based on the idea, supported by modern 
many-body calculations, of a continuously varying degree of localization
of the $4f$ electrons as a function of compression. The functional forms
used are motivated by many-body calculations, with parameters determined
by experiments, resulting in a physics-based empirical EOS.
In the large-volume $\gamma$ phase, degeneracy of the 
$j = 5/2$ state of the localized $4f$ electron makes a large contribution
to the entropy. Rapid variation of the thermal electronic free energy
with volume leads to unusually large electronic contributions to the
pressure and thermodynamic Gr\"{u}neisen parameter. The static lattice
energy has two local minima, with the $\alpha$-like minimum lower
than the $\gamma$-like one by 6.3 meV/atom.
\end{abstract}

% Head 1
\section{INTRODUCTION}

\begin{figure}
  \includegraphics[width=.5\textwidth]{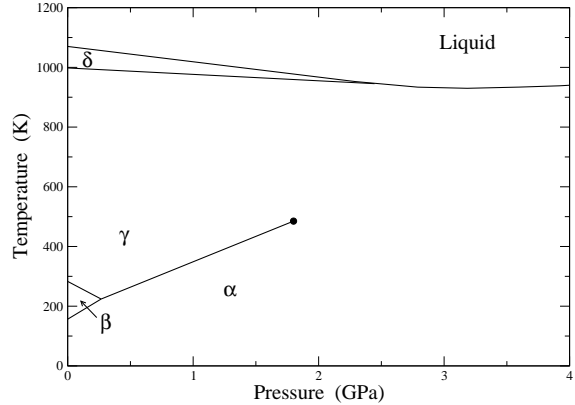}
 \caption{Phase diagram of Ce at low pressure 
\cite{lipp2008kondo} and \cite{schiwek2002phases}. The $\gamma$
and $\alpha$ phases have the fcc structure. Their phase boundary ends at a critical point marked by the dot. The $\beta$ phase has the dhcp structure,
and $\delta$ is bcc.}
 \label{lowp_phase_diag}
\end{figure}

The element Ce exhibits many interesting phenomena. Chief among these is the
isostructural $\gamma-\alpha$ phase transition. 
This transition involves a discontinuity in volume as a function of pressure,
but no change of crystal structure, both phases being fcc. The volume 
discontinuity decreases with increasing temperature, and 
vanishes at a critical point at
$T=480$~K and $P=1.5$~GPa. Having a volume change but no symmetry change,
and ending in a critical point, the transition is analogous to the liquid-vapor
transition.

The $\gamma-\alpha$ transition has large volume and entropy changes of
15\% and $1.5 k_B$/atom, respectively, at room 
temperature \cite{lipp2008kondo,amadon2006entropy}. The phase diagram in this
region is shown in Fig. \ref{lowp_phase_diag}. In addition to the
fcc $\gamma$ and $\alpha$ phases, at ambient pressure and just below 
room temperature, there is the dhcp structured $\beta$ phase. At high
temperature, there is a bcc $\delta$ phase. The melting curve is
anomalous with $dT_{\rm melt}/dP < 0$ at low pressure. The melting temperature
reaches a minimum near 3.2 GPa, and increases at higher pressures.
Other phases enter beyond the range of the plot. These show large
history dependence and other anomalies, resulting in significant uncertainty 
regarding the equilibrium phase diagram at 
high pressures \cite{schiwek2002phases,tsiok2001}.
Under dynamic compression, the large volume collapse of the $\gamma-\alpha$
transition leads to strong shock heating, contributing to the 
unusually low shock melting pressure of 10 GPa \cite{jensen2010shock}. In addition, the bulk
modulus of the $\gamma$ phase decreases on compression, leading
to anomalous spreading of compression waves \cite{jensen2012dynamic}.

The existence of the isostructural transition in Ce has its origins in the
behavior of its $4f$ electrons. Ce belongs to the lanthanide series where
the $4f$ shell is partially full, and it has a nominal configuration of 
$[{\rm Xe}] 4f^1 5d^1 6s^2$. The $4f$ atomic state is energetically 
in the valence, but spatially localized. This results in narrow energy bands
in the solid, so that the on-site Coulomb repulsion is larger than
the bandwidth. Electron correlation effects are significant, and the
mean-field picture of Bloch band states breaks down. Compression increases 
the bandwidth, effectively reducing the strength of correlations. In the strong 
interaction limit, the $4f$ electrons are localized on atomic sites. 
We show below that the localized $4f$ limit gives a good description of 
the thermodynamic properties of the large-volume $\gamma$ phase. The 
$\gamma-\alpha$ transition is associated with a change from localized
to itinerant behavior of the $4f$ electrons.

The earliest equation of state model for the transition is due to
Aptekar and Ponyatovskiy \cite{aptekar1968theory}, who considered 
the transition to be due to
a change of valence of Ce ions. They constructed a pseudo-alloy 
model, picturing two types of Ce atoms, with a concentration-dependent
configuration entropy,
\begin{equation}
S_{\rm conf} = N k_B \left[ x \ln(x) + (1-x)\ln(1-x)\right]
\end{equation}
 corresponding to the number of arrangements
of the two types of atoms, where $x$ is the 
concentration of $\alpha$-like atoms. 
Later, Johansson {\em et al.} \cite{johansson1995phasediag} formulated a 
model based on the picture
of a Mott transition between localized and de-localized $4f$ electrons.
They incorporated a magnetic entropy in the localized, $\gamma$-like 
state, which is included in our model as well. 
Similar to Aptekar and Ponyatovskiy, they adopt a pseudo-alloy
approach, with the same configuration entropy. 
More recently, Elkin {\em et al.} have developed an equation of state
for several phases of Ce \cite{elkin2011eos}, which describes a wide
range of data well.  Their treatment 
of the $\gamma-\alpha$ transition also invokes a pseudo-alloy picture,
similar to that of Aptekar and Ponyatovskiy. They consider the
two atom types in the pseudo-alloy to be in pressure equilibrium.

The present model is motivated by both practical and theoretical
considerations. We wish to have a parameterized equation of state for
the $\gamma-\alpha$ region that is compatible with our existing framework
for generating wide-ranging multi-phase equations of 
state \cite{chisolm2005multi}.
We would like to be able to use the analytic formulation of the model 
in-line in hydrodynamic simulations which include phase transition 
kinetics \cite{greeff_dyn_phase_2016}.
The model of Elkin {et al.} \cite{elkin2011eos} is difficult to use in this
setting because of the requirement of pressure equilibrium, which introduces
an additional layer of transformation between volume and pressure as the
independent variable.
Theoretically, the pseudo-alloy picture is dubious. There is no 
way to assign an individual atom as localized or itinerant, and the meaning
of the configuration entropy $S_{\rm conf}$ is unclear.
Here we formulate a model for the Helmholtz free energy of fcc Ce in the form
of a sum of static lattice energy and ionic and electronic excitation
contributions,
\begin{equation}
F(V,T) = \phi(V) + F_{\rm ion}(V,T) + F_{\rm el}(V,T)  \, .
\end{equation}
Both the static lattice energy $\phi$ and the electronic excitation terms have
unusual functional forms related to the de-localization of the $4f$ electrons
under compression. We use modern many-body calculations to guide the functional
forms of the models \cite{mcmahan2003ce_thermo_dmft,bieder2014ce_thermo_dmft,tian2015lda+gutz}. These show a rapid, but continuous increase in the 
$4f$ spectral weight at the Fermi surface, and the strength of the
$4f$ quasiparticle pole under compression. Our model represents this
with a continuous transition from a localized to Fermi-liquid behavior. 
The many-body calculations also show a non-convex
variation of $\phi(V)$, and our model incorporates this.

\section{EOS MODEL}

\subsection{Electronic Excitation}

\begin{figure}
   \includegraphics[width=.5\textwidth]{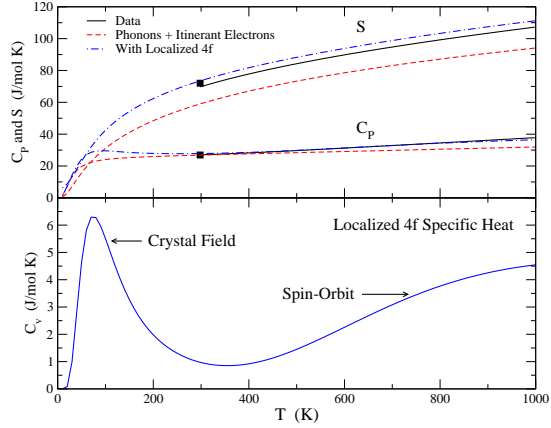}
   \caption{Entropy and specific heat of $\gamma$ Ce. Solid black curve
is data from Konings and Bene\v{s} \cite{konings2010thermo} 
and black squares are from  Gschneidner \cite{gschneidner1990}.
Red dashed curves are phonon and itinerant electron contributions  only, blue 
dot-dash curves also include localized $4f$ electrons.
Lower panel shows the contribution of the localized $4f$ electrons
to the specific heat.}
   \label{f_loc_thermal}
\end{figure}
\begin{figure}
   \includegraphics[width=.4\textwidth]{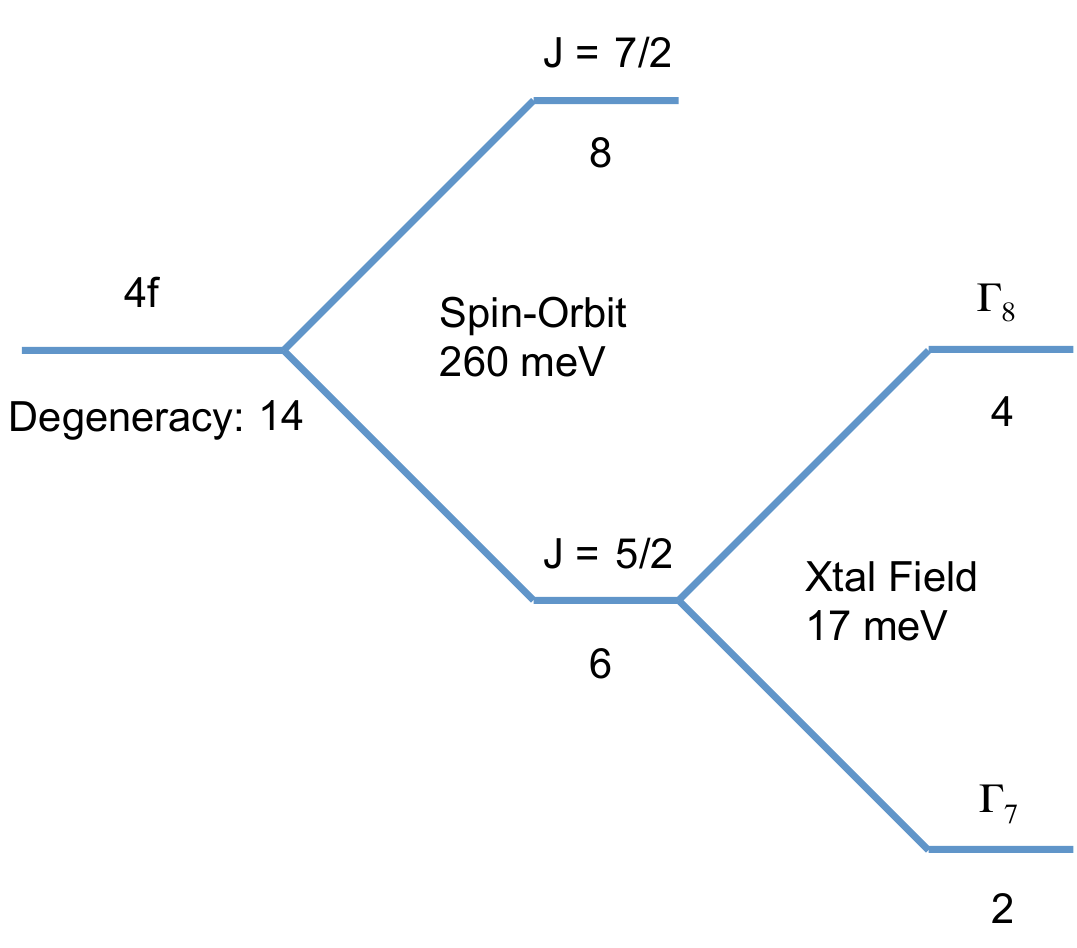}
   \caption{Energy levels of a $4f$ electron in $\gamma$ 
         Ce \cite{manley2010entropy,arajs1962spec_ht}.
             Numbers below horizontal lines give degeneracies.}
   \label{f_loc_levels}
\end{figure}

Standard equation of state models for 
metals include the phonon contributions through
an effective Debye model, and an electronic term of the form $F_{\rm el}
= - \Gamma T^2$, where $\Gamma$ is the Sommerfeld coefficient, which is 
proportional to the electronic density of states \cite{greeff_gold_eos}. 
Good estimates for these parameters are available from measurements for
$\gamma$ Ce. The phonon dispersion curves were measured by Stassis {\em et al.}
\cite{stassis1979phonons}. From these data, the logarithmic moment,
 $\theta_0 = h \nu_0/ k_B$, which determines the entropy 
in the high temperature limit,
is 116 K. Using this as the effective Debye temperature, and the $\gamma$
phase Sommerfeld coefficient 
$\Gamma = 6.2$~mJ/mol K$^2$ \cite{manley2010entropy}, we obtain 
the red dashed curves in the upper panel of Fig. \ref{f_loc_thermal}. 
It can be seen that the entropy obtained from these parameters is substantially
smaller than the experimental entropy \cite{gschneidner1990,konings2010thermo}.
The difference is 12 J/mol K = 1.4 $k_B/$atom, which is similar to  the
$\gamma-\alpha$ entropy change. The entropy can be increased by raising
$\Gamma$, but that also increases the specific heat, in conflict with data.
The only way to increase the entropy within this model, without increasing the
high temperature specific heat, is to decrease the Debye temperature. To
obtain agreement with data,  we must set $\theta_D$ to 80 K. A range of
115-137 K is given for various measurements aggregated by Lipp {\em et al.}
\cite{lipp2013debye}, so 80 K is well outside the range of experimental data.

To account for the measured entropy of $\gamma$ Ce, we must include the
effects of the localized $4f$ electron. The energy levels of the
$\gamma$ Ce $4f$ electron are shown in Fig. \ref{f_loc_levels}. 
Starting from a 14-fold orbital and spin degeneracy, spin orbit
coupling splits these into a 6-fold degenerate $j=5/2$ state and
an 8-fold degenerate $j=7/2$ state, with the latter lying 260 meV
above the former \cite{arajs1962spec_ht}. Crystal fields further
split the $j=5/2$ state into 2-fold degenerate $\Gamma_7$ 
and 4-fold $\Gamma_8$ representations, which are separated by 
17 meV \cite{manley2010entropy}. Spins on different atomic sites are 
correlated at a still lower energy scale. For instance $\beta$ Ce
undergoes a magnetic ordering transition at $\sim 12$ K, with a 
corresponding energy scale $\sim 1$~meV \cite{lock57susc}. Since we 
are concerned with properties near room temperature and above, these
energies are negligible, and we can regard the atomic moments as independent.
Then the thermodynamic functions associated with the localized $4f$
electrons are obtained straightforwardly from the partition function.
The Helmholtz free energy is $F_{\rm loc} = -N k_B T \ln Z_{\rm loc}$,
where $Z_{\rm loc} = \sum_i g_i \exp( -\beta \epsilon_i)$, and
$ g_i$ and $\epsilon_i$ are the degeneracies  and energies in Fig.
\ref{f_loc_levels}.

Including this contribution leads to the blue dash-dot curves in Fig.
\ref{f_loc_thermal}, which are in much better agreement with experiment.
The lower panel shows the contribution of $F_{\rm loc}$ to the specific
heat. The crystal field splitting leads to a peak at low temperature,
below the stability of the $\gamma$ phase. In the temperature range of 
interest, essentially from room temperature up, the main effect
of the localized $4f$ electrons is to increase the entropy by
$N k_B \ln(6)$, where 6 is the combined number of $j = 5/2$ states.
Excitations to the $j = 7/2$ states contribute an enhancement of the
high temperature specific heat, which also improves agreement with data.
The entropy from our complete model, described below, is in somewhat better
agreement with data because the $4f$ electrons are not completely localized
at the ambient $\gamma$ phase volume.

\begin{figure}
  \includegraphics[width=.5\textwidth]{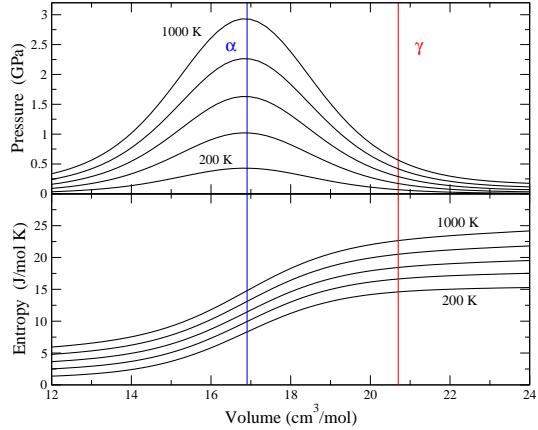}
 \caption{Entropy and pressure from the electron free energy
   $F_{\rm el}$ given in Eq. (\ref{F_el}).}
 \label{s_and_p_el}
\end{figure}

To complete the electronic free energy model, we must incorporate a volume 
dependence such that under compression, the free energy changes over to 
that of a normal Fermi liquid. Our model is motivated by DMFT calculations 
that show a smooth buildup of $4f$ spectral weight at the Fermi 
surface \cite{mcmahan2003ce_thermo_dmft,bieder2014ce_thermo_dmft}.
The entropy curves 
shown in Fig. 2 of Bieder and Amadon \cite{bieder2014ce_thermo_dmft},
show a component at large volume that is nearly temperature independent
above 300 K, and is close to $S/Nk_B = \ln(14)$. Their calculations 
do not include 
spin-orbit coupling, so this corresponds to the $\ln(6)$ in our model
associated with the local $4f$ moment. At small volumes, their entropy 
is close to linear in $T$, as expected for a normal Fermi liquid.
We therefore formulate our electron thermal model for the $4f$
electrons as,
\begin{equation}
F_{f}(V,T) = h(V) F_{\rm loc}(T) 
               - \left[1 - h(V)\right] \frac{1}{2} \Gamma_f T^2
\label{F_f}
\end{equation}
where $h$ is a switch function going from one at large volume to
zero at small volume.
\begin{equation}
h(V) = e^x/(e^x + 1)
\label{locfn}
\end{equation}
where $x=(V-V_{\rm loc})/\Delta V_{\rm loc}$. The parameters $V_{\rm loc}$ and
$\Delta V_{\rm loc}$ control the location and width of the transition, 
respectively.
We have chosen to make $\Gamma_f$ volume independent and to make
$h$ a function of $V$ alone in order to limit the number of parameters
in the model. Initial values of the parameters $V_{\rm loc}$ and 
$\Delta V_{\rm loc}$ were chosen by matching the 600 K isotherm of the DMFT
calculations in
Fig. 2 of Bieder and Amadon \cite{bieder2014ce_thermo_dmft}. These
were then fine-tuned to match data on the phase diagram. The assumption
that $h$ depends only on $V$ is somewhat limiting. Matching the phase diagram 
tends to favor smaller values of $V_{\rm loc}$, which correspond to 
higher temperatures in the DMFT calculations.

The complete electron model includes the other conduction electrons
of $s$, $p$, and $d$-like character. These are represented together 
with a separate volume-dependent Sommerfeld coefficient,
\begin{equation}
F_{\rm el}(V,T) = F_{f}(V,T)
               - \frac{1}{2} \Gamma_{spd}(V) T^2
\label{F_el}
\end{equation}
where the volume dependence is given by,
$\Gamma_{spd}(V) = \Gamma_{0\, spd} (V/V_{0\, spd})^\kappa$.

\newpage

In addition to the energy levels and degeneracies listed in Fig.
\ref{f_loc_levels} for the localized $4f$ electrons, the following
parameters are used,
\begin{eqnarray}
\Gamma_{0\, spd} & = & 6.12 \, {\rm mJ/mol \, K^2} \nonumber \\
\kappa & = &  1.0 \nonumber \\
\Gamma_f & = & 2.13 \, {\rm mJ/mol \, K^2}  \nonumber \\
V_{\rm loc} & = & 16.85 \, {\rm cm}^3/{\rm mol}  \nonumber \\
\Delta V_{\rm loc} & = & 1.20 \, {\rm cm}^3/{\rm mol}
\end{eqnarray}

Figure \ref{s_and_p_el} shows the entropy and pressure corresponding to the electronic
free energy, Eq. (\ref{F_el}).  At large volumes, the entropy rises quickly to
the $N k_B \ln(6)$ value from the local moments. The transition volume
from localized to itinerant $4f$ electrons is close to the ambient pressure
$\alpha$ phase volume. At small volumes, the entropy is linear in $T$,
with coefficient $\Gamma_f + \Gamma_{spd}$. The rapid variation of the
entropy with volume is accompanied by a peak in the electronic contribution 
to the pressure, in accord with the thermodynamic identity 
$\left(\partial S/\partial V\right)_T = \left(\partial P/\partial T\right)_V$.
The electron-thermal pressure is exceptionally large in the transition region,
and is comparable to the critical pressure. The thermodynamic Gr\"uneisen
parameter $\gamma_{\rm th} = V \left(\partial P/\partial E\right)_V$ is observed
to show a peak for volumes near the $\gamma-\alpha$ phase transition 
\cite{lipp2008kondo}. In the present model,  this peak is attributed to
thermal electronic excitations, rather than lattice vibrations.

\subsection{Static Lattice Energy}

\begin{figure}
  \includegraphics[width=.5\textwidth]{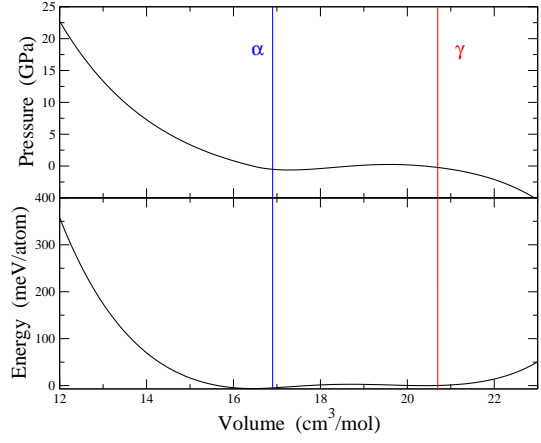}
 \caption{Static lattice energy and corresponding pressure.}
 \label{e_and_p_static}
\end{figure}

Our static lattice energy is modeled on results from many-body calculations
that show non-convex shape of $\phi(V)$, leading to double tangents.
For instance, Fig. 1 of Tian {\em et al.} \cite{tian2015lda+gutz} shows 
results of LDA + Gutzwiller calculations on  $\phi(V)$ for different values of
the on-site coulomb interaction parameter $U$. These show ``normal"
regions with $e^2\phi/d V^2 >0$ separated by an anomalous region with
negative curvature.

We therefore formulate our static lattice energy as follows. We define 
a $\gamma$-like region at larger volumes, and an $\alpha$-like 
one at small volumes,
each of which is described by the Vinet form.
\begin{eqnarray}
\phi(V) & = & \phi_* +\frac{4 V_* B_*}{(B'_* - 1)^2} 
            \left[1 -(1+X) e^{-X}\right]
      \\
 X & = & \frac{3}{2} (B'_* -1) \left[ \left(V/V_*\right)^{1/3} -1 \right]
\end{eqnarray}
The quantities $\phi_*$, $V_*$, $B_*$, and $B'_*$ are the  energy, volume,
bulk modulus, and its pressure derivative evaluated at the minimum of
the corresponding segment. The ambient density and bulk modulus are primarily
controlled by the parameters of the $\gamma$-like Vinet curve. The
$\alpha$-like Vinet curve controls the cold properties on the high
pressure side of the transition. It gives the dominant contribution
to the room temperature isotherm at high pressure. We have used room 
temperature data to 208 GPa \cite{vohra1999ultrapress} to determine 
parameters for the $\alpha$-like Vinet curve. This includes the $\alpha''$
and bct phases as well as $\alpha$, so our high pressure $\phi$ incorporates
these phases in a smoothed way.

We specify the static lattice  transition pressure $P_{\rm tr}$ as
an input. 
This is approximately
the $T=0$ transition pressure, but because the zero-point vibration energy is
included in $F_{\rm ion}$, the two are not exactly the same. 
We solve for the volumes $V_1$ and $V_2$ where the pressure is equal to
$P_{\rm tr}$ on the $\gamma$ and $\alpha$-like Vinet curves, respectively.
For $V > V_1$, we use the $\gamma$-like Vinet curve, and for $V < V_2$
the $\alpha$-like curve. In between, we use a polynomial in $V$. The
coefficients are chosen so that there is continuity of the pressure
and its volume derivative at $V_1$ and $V_2$, and so that the
enthalpies $H = E +PV$ are equal at these two volumes. This is the
condition of phase equilibrium. These conditions require a fifth order 
polynomial. The current set of parameters are,
\begin{eqnarray}
V_*^{\gamma} & = & 20.4 \, {\rm cm}^3/{\rm mol} \nonumber \\
B_*^{\gamma} & = & 13.0 \, {\rm GPa}  \nonumber \\
B_*^{\prime\gamma} & = & -20 \,  \nonumber \\
V_*^{\alpha} & = & 16.54 \, {\rm cm}^3/{\rm mol} \nonumber \\
B_*^{\alpha} & = & 27.0 \, {\rm GPa}  \nonumber \\
B_*^{\prime\alpha} & = & 6.5  \nonumber \\
P_{\rm tr} & = & -0.15 \, {\rm GPa}
\end{eqnarray}

The resulting $\phi(V)$ and its corresponding pressure are shown in Fig. 
\ref{e_and_p_static}. The $\gamma$ phase minimum is higher than the 
$\alpha$ phase by 6.3 meV/atom. The local maximum between the two 
is 2.9 meV/atom above the $\gamma$ minimum. This conveys a sense of the
small energy scales at work in the phase transition. Our empirical
energy curve is similar to Tian's \cite{tian2015lda+gutz} result
for $U=4.5$~eV. However, using
other criteria, they selected $U=4.0$~eV as the best value, which gives a much
larger energy difference between $\alpha$ and $\gamma$. 
This indicates that some fine-tuning of $U$ is
likely to be needed to get a qualitatively correct phase diagram in
correlated electron calculations.

\subsection{Ion Motion}

\begin{figure}
  \includegraphics[width=.5\textwidth]{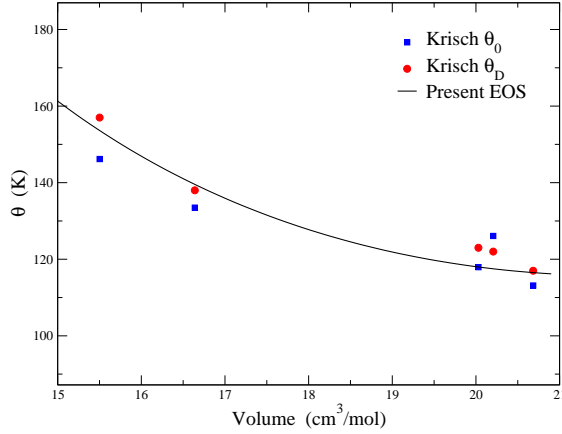}
 \caption{Effective Debye temperatures as functions of the volume.
          Squares are $\theta_0$ determined by integrating the
          density of states from Krisch {\em et al.} \cite{krisch2011phonons}
           and circles are their reported values based on Debye-Waller 
          factors. Solid black curve is the present model.}
 \label{theta_of_v}
\end{figure}

The ion motion free energy is simply the Debye model with a volume-dependent
effective Debye temperature $\theta(V)$. The dependence of $\theta$ on $V$
is prescribed through the phonon Gr\"{u}neisen parameter
$\gamma^{\rm ph} = - d \ln \theta / d \ln V$
whose volume dependence we take to be,
\begin{equation}
\gamma(V) = \gamma^{\infty} + A \left( \frac{V}{V_0}\right) 
                + B \left( \frac{V}{V_0}\right)^2 \, ,
\label{gamma_of_v}
\end{equation}
where $\gamma^{\infty}$, $A$, and $B$ are material parameters. We specify
$A$ and $B$ by giving $\gamma(V_0)$ and $q=d \ln(\gamma) /d \ln(V)|_{V_0}$.

Krisch {\em et al.} present measurements of the phonon dispersion curves
under pressure. From their Debye-Waller factors they inferred the
volume dependence of an effective Debye temperature $\theta_D(V)$. 
They also give graphs of the density of states $g(\omega)$, from which we could
determine moments. In particular, the logarithmic moment
\begin{equation}
\omega_0  =  e^{1/3} \exp\left[ \int d\omega \, g(\omega) \ln\omega \right ]
\label{log_moment}
\end{equation}
is of interest because it determines the entropy in the high temperature 
(classical) limit, and $\theta_0 = \hbar \omega_0/k_B$ is often used
as an effective Debye temperature.
Figure \ref{theta_of_v} shows these two effective Debye temperatures
as functions of volume, along with the present model,
which follows equation \ref{gamma_of_v}. The two moments follow generally 
the same trends and our model is between them. The experimental moments
are not smooth enough to usefully determine $\gamma_{\rm ph} = d \ln(\theta)/
d \ln(V)$ by numerical differentiation. Some general features are supported
by the data. The phonon Gr\"{u}neisen parameter $\gamma_{\rm ph}$ is small
at ambient conditions, and through the transition region. It 
increases under compression, in contrast to the usual trend.
Our model uses the following parameters for the vibrational free energy,
\begin{eqnarray}
V_0 & = & 20.67 \, {\rm cm}^3/{\rm mol}  \nonumber \\
\theta(V_0) & = & 116.5 \, {\rm K} \nonumber \\
\gamma(V_0) & = &  0.3 \nonumber \\
q & = & -20 \nonumber \\
\gamma^{\infty} & = & 2/3
\end{eqnarray}
The large negative value of $q$ is unusual, reflecting the anomalous 
increase of $\gamma_{\rm ph}$ under compression.

\section{RESULTS}

\begin{figure}
  \includegraphics[width=.5\textwidth]{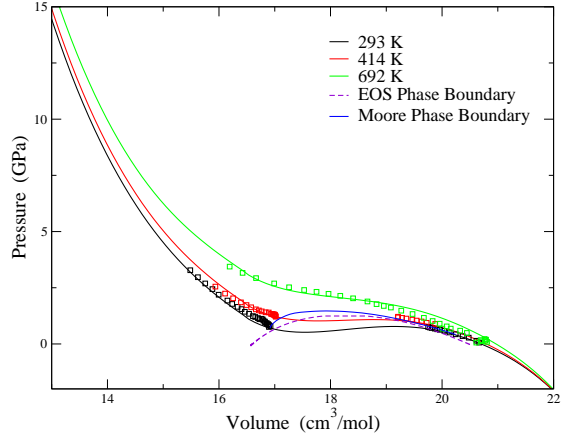}
 \caption{Isotherms and $\gamma-\alpha$ phase boundary. Symbols are data 
from Lipp {\em et al.} \cite{lipp2008kondo}. Solid blue phase 
boundary is from Moore {\em et al.} \cite{moore2011collapse}.}
 \label{isotherms}
\end{figure}

\begin{figure}
  \includegraphics[width=.5\textwidth]{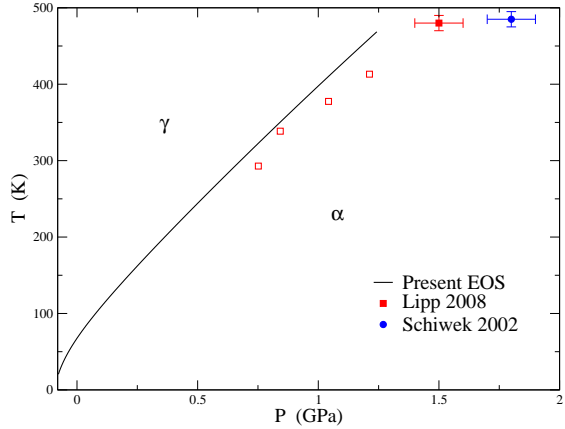}
 \caption{The $\gamma-\alpha$ phase boundary with 
data \cite{lipp2008kondo,schiwek2002phases} Filled symbols are measurements
of the critical point.}
 \label{gamma_alpha_bndry}
\end{figure}

Figure \ref{isotherms} shows some isotherms in the $\gamma-\alpha$ transition
region, along with the phase boundary. 
The phase boundary for the EOS is obtained by solving 
$P(V_\gamma,T) = P(V_\alpha,T) $ simultaneously with 
$G(V_\gamma,T) = G(V_\alpha,T) $ where $P=-\partial F/ \partial V$ is the 
pressure, and $G = F +PV$ is the Gibbs free energy. The $P(V)$ compression 
data is obtained in diamond anvil cell experiments by Lipp {\em et al.} 
\cite{lipp2008kondo}. The experimental phase boundary is from 
Moore {\em et al.} \cite{moore2011collapse}.
The isotherms are generally matched well by the model, both their 
volume and temperature dependence. 
Figure \ref{gamma_alpha_bndry} shows the $\gamma-\alpha$ phase boundary
in the $T,P$ plane with data \cite{schiwek2002phases,lipp2008kondo}.
The model's critical temperature is in good agreement with data. The
phase boundary is in overall good agreement with data, both in the
$P,V$ and $T,P$ planes, except that the 
transition pressures are too low by about 0.2 GPa.

\begin{figure}
  \includegraphics[width=.5\textwidth]{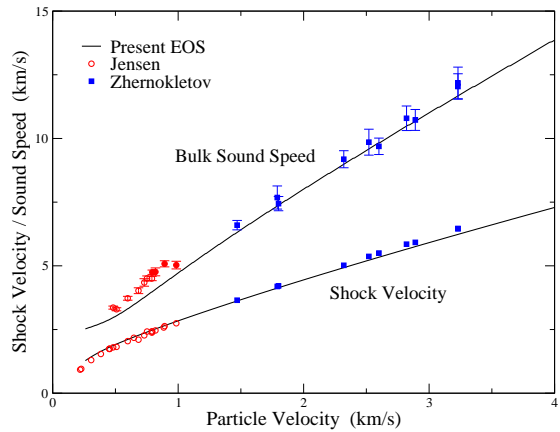}
 \caption{Hugoniot and bulk sound speed of shocked Ce. Data from
          Jensen and Cherne \cite{jensen2012dynamic} and 
          Zhernokletov {\em et al.} \cite{zhernokletov2008soundspeed}.}
 \label{usup_and_cb}
\end{figure}

Figure \ref{usup_and_cb} shows the Hugoniot and
bulk sound speed from the model compared with 
data \cite{jensen2012dynamic,zhernokletov2008soundspeed}. While the model
refers to the solid $\gamma-\alpha$ phases, the data above $U_p \sim 0.8$~km/s
is in the liquid. In spite of this, the high pressure Hugoniot is in 
fair agreement with the data. Because we have used data from high 
pressure phase to parameterize our $\alpha$-like static lattice pressure,
this agreement shows that the EOS of the liquid is similar to that
of the high pressure solid phases. Inclusion of a distinct liquid phase should
move the Hugoniot higher, in the right direction to improve agreement with data.
The sound speeds at low pressure are systematically low. The peak of
the sound speed near melting was not reproduced with any values of the parameters we tried.

\section{CONCLUSIONS}

We have formulated a model for the free energy of fcc Ce in the 
region of the $\alpha-\gamma$ isostructural phase transition. The
pseudo-alloy 
picture \cite{aptekar1968theory,johansson1995phasediag,elkin2011eos} is not 
invoked, and instead the $4f$ electrons are considered to transition rapidly 
but smoothly from localized to de-localized under compression.
This de-localization is represented by a non-covex static lattice energy
and an unusual form for the electronic excitation free energy.
Electronic  excitations account for almost all of the entropy
change at the phase boundary, and contribute substantially to the
pressure, owing to the rapid variation of the electronic spectrum with volume.

The model  has correct qualitative behavior and describes a variety of 
data rather well. Improvements should
be possible by further refinement of the parameters. The present parameters
were obtained from estimates based on data and theoretical calculations,
 which were then partially optimized by hand. Automatic optimization 
will be needed to systematically improve them. A potentially significant 
physical 
improvement to the model would be to generalize 
the transition from localized to de-localized
$4f$ electrons, as governed by
equations \ref{F_f} and \ref{locfn}, to depend on volume
and temperature, rather than volume alone. Entropy curves 
from microscopic
calculations indicate \cite{mcmahan2003ce_thermo_dmft,bieder2014ce_thermo_dmft}
the need for explicit temperature dependence.
A tractable way to do this might be to follow the method of reference
\cite{lipp2008kondo} by introducing a 
scaled temperature, $h(V,T) = f(T/T_K(V))$,
where $T_K$ is the Kondo temperature.

\section{ACKNOWLEDGMENTS}

We gratefully acknowledge support from Advanced
Simulation and Computing at LANL which is operated
by LANS, LLC for the NNSA of the U.S. DOE under Contract
No. DE-AC52-06NA25396, and helpful discussions with John Wills 
and Frank Cherne.

% References

%\nocite{*}
\bibliographystyle{aipnum-cp}%
\bibliography{ce}%

\end{document}